\documentclass[pre,floatfix,twocolumn, 10pt]{revtex4} 
\usepackage{graphicx}
\usepackage{amssymb}
\usepackage{natbib}
\usepackage{dcolumn}
\usepackage{bm}
\usepackage[normalem]{ulem} 
\usepackage{color}
\usepackage[colorlinks=true, linkcolor=blue, citecolor=blue]{hyperref}
\usepackage{subfigure}
\usepackage{graphicx,psfrag,xspace}
\usepackage{color}
\usepackage{amssymb,amsfonts,amsmath}
\usepackage{setspace}

\begin{document}

\title{Fatigue crack growth without plasticity under cyclically reversed shear}


\author{E. A. Jagla}
\affiliation{Centro At\'omico Bariloche, Instituto Balseiro,
Comisi\'on Nacional de Energ\'ia At\'omica\\ CNEA, CONICET, UNCUYO,
Av.~Bustillo 9500 (R8402AGP) Bariloche, R\'io Negro, Argentina}

\begin{abstract}
%
We show that cyclic fatigue can occur in a system whose elements behave elastically up to an irreversible failure threshold, without plasticity, friction, or any time-dependent process. The model represents a weak interface between two elastic bodies loaded in shear. Under sufficiently large cyclic loading below the static failure threshold, a pre-existing flaw advances progressively, by a few lattice spacings per cycle. Two conditions are needed for this behavior: each bond fails more easily under one sense of the shear than the opposite, 
and the shear must be reversed during the cycle by an amount large enough to break bonds that have not yet failed.
The model thus identifies the minimal conditions for an intrinsic, plasticity-free fatigue mechanism. Such conditions may be met at interfaces and weak planes subjected to strongly reversed shear, as in fretting, rolling contact, and reversed torsion, or in cyclic shear of cemented granular materials.

\end{abstract}
\maketitle

\section{Introduction}

A defining characteristic of cyclic fatigue is the subcritical and
progressive accumulation of damage, where each loading cycle contributes
only an incremental fraction to the total degradation of the
material~\cite{suresh1998}. Since the external load returns to the same
value at the end of every cycle, some irreversible process must prevent
successive cycles from simply retracing the same sequence of states.
Identifying this process at the microscopic level is a central question
in the physics of fatigue.

In metals and alloys, cyclic fatigue is intimately linked to plasticity\cite{suresh1998}.
Under repeated loading, dislocation motion organizes into persistent slip bands, whose
imperfect reversibility produces surface extrusions and intrusions that act as crack
nucleation sites\cite{thompson1956,essmann1981}. Microstructurally small cracks then grow
along the active slip planes, under predominantly shear (Mode-II or
Mode-III) conditions\cite{forsyth1961,miller1982,suresh1984propagation}, and eventually
turn into Mode-I cracks that advance perpendicular to the maximum tensile stress. 
The crack advance per cycle results from the irreversible plastic blunting and
re-sharpening of the crack tip during each loading-unloading sequence, which leaves the
characteristic striations on the fracture surfaces\cite{laird1967pathogenesis}, while the
plastic wake left behind the advancing crack modifies the effective driving force through
crack closure\cite{elber1971significance}. At all these stages, the irreversibility of
dislocation motion is the ultimate source of the cycle-by-cycle accumulation of
damage\cite{mughrabi2009}.

Ceramics and other brittle solids are examples of materials where
plasticity is essentially absent at ambient temperature, yet
many structural ceramics display genuine cyclic fatigue, namely crack growth
under cyclic loading at stress intensities well below those required for
crack growth under monotonic or sustained loads
\cite{Dauskardt1987,Dauskardt1992,Gilbert1995a,Gilbert1995b,Gilbert1996}.
In the standard rationalization of these
observations~\cite{Ritchie1988,Ritchie1999,Ritchie2000}, the mechanism of
crack-tip advance is the same as under monotonic loading, and the effect of
cycling is extrinsic: in grain-bridging materials such as alumina, silicon
nitride, or silicon carbide, repeated unloading and reloading progressively
degrades the bridges in the crack wake by frictional wear, reducing the
shielding that provides the toughness of the
material~\cite{Lathabai1991,Gilbert1995b,Dauskardt1993}. 
Consistently with
this picture, fatigue crack growth rates in ceramics depend much more strongly
on the maximum stress intensity $K_{\max}$ than on $\Delta K$, and the
materials most toughened by bridging tend to be the most susceptible to cyclic
degradation \cite{Ritchie1999,Ritchie2000}. This trade-off between toughness
and fatigue resistance remains relevant in the context of current strategies
for the microstructural toughening of ceramics \cite{GavaldaDiaz2025}.

A similar picture emerges from statistical models of fracture. Most models
of fatigue in disordered media introduce an explicit, phenomenological
damage-accumulation rule, in which each element accumulates damage as a
function of time or of the number of cycles, typically as a power of its
local load~\cite{SornetteVanneste1992,Kun2007,Carmona2007,Kun2008,Vieira2008,Oliveira2012,Araujo2016}.
Cohesive-zone descriptions of fatigue based on a damage-evolution law follow the
same logic \cite{Roe2003}, and subcritical growth can also result from thermally
activated rupture \cite{Roux2000,Santucci2004}. 
In all these cases the cyclic
nature of the load is not essential, and the same ingredients produce delayed
failure under a constant load.

In strictly rate-independent models, where damage
accumulates only because the load is cycled, some form of irreversibility
must operate before the elements fail: plastic
deformation\cite{GuozdenJagla2012,Fiocco2013,Leishangthem2017,Maity2026,ferrero2025cyclic},
hysteresis of the cohesive law \cite{Nguyen2001}, or frictional sliding
\cite{Lathabai1991,AlonsoMarroquin2004}. 
The reason is general. If the elements
respond elastically until failure, and the load is varied
proportionally between two values of the same sign, then after the first
cycle the system returns to the same equilibrium configuration every time
the maximum load is reached, no new element reaches its threshold, and
damage stops: the system shakes down to a purely elastic
response~\cite{Nguyen2001,GuozdenJagla2012}. Consistently, the fatigue
threshold of nearly ideal polymer networks, in which dissipation is
negligible, is found to be close to their intrinsic fracture
energy~\cite{Zheng2022}.

Here we show that this limitation can be circumvented without plasticity, friction,
or time dependence: when each elastic element breaks more easily under one
orientation of the applied shear and the loading direction is reversed, the two
halves of each cycle bring different subsets of bonds to failure, and damage
can accumulate cycle after cycle.

The natural domain of this kind of mechanism
is that of interfaces and weak planes subjected to
strongly reversed shear. Examples are rolling contacts, where the
subsurface orthogonal shear stress is fully reversed as the load passes
over a given point, and the compressive normal stress suppresses crack
kinking, so that subsurface cracks grow in
Mode~II~\cite{Johnson1985,Fujii2001,Kida2012}; fretting contacts under
oscillating tangential loads~\cite{HillsNowell1994}; components under
reversed torsion, for which slow Mode~II fatigue crack growth has been
reported in silicon nitride~\cite{Kida2012}; and cemented granular
materials and rock joints under cyclic
shear~\cite{Potyondy2004,Cerfontaine2018}. Frictional interfaces, whose
transition to sliding is itself mediated by shear
cracks~\cite{Svetlizky2014}, are a closely related case. In all of them,
plastic deformation may be negligible and the shear acting on the relevant
plane reverses during each cycle.

The rest of the paper is organized as follows. In Sec.~\ref{sec:model} we
define the model and the loading protocol. Section~\ref{sec:results}
presents the results for a straight seed crack, including the dependence
of the advance per cycle on the loading amplitude and on the load ratio.
In Sec.~\ref{sec:geometry} we extend the results to circular cracks of different 
radii. Section~\ref{sec:discussion}
discusses the results and compares them with recent studies of fatigue
in amorphous solids under cyclic
shear~\cite{Maity2026,Maity2026b,Khandare2026,Bhaumik2026}. In the Supplementary Material
we provide an example of the proposed mechanism acting on a full three-dimensional system of interacting particles.

\section{Model Formulation}
~\label{sec:model}
\begin{figure}[h]
\includegraphics[width=6cm,clip=true]{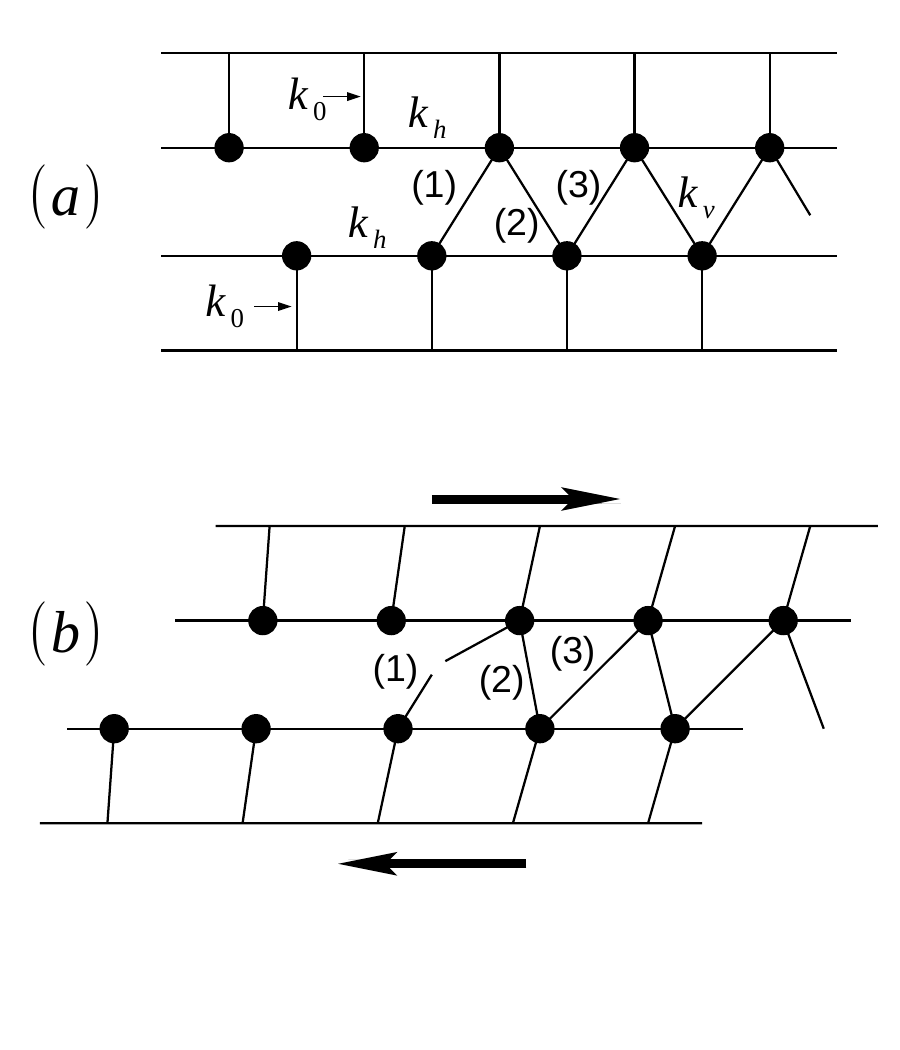}
\caption{A one-dimensional sketch of the model. In (a) the system is in its minimum energy configuration in the absence of applied external strain. Notice that a few inter-layer springs are assumed to be broken, and therefore are not drawn. (b) The configuration when a strain $\gamma$ is applied, as indicated by the arrows. Odd-numbered springs are {\em stretched}, while even-numbered springs are {\em compressed}. Among the former, spring (1) is the most heavily stretched due to stress concentration effect caused by previously broken springs, and will be the first to break, as indicated. Notice that after spring (1) fails, (2) can prevent (3) from failing. In these idealized conditions, one single spring is broken on each reversion of the applied strain.
}
\label{sketch}
\end{figure}

We consider two triangular lattices of particles in their compact arrangement as they appear for instance in fcc, or hcp structures. 
Particles interact with their nearest neighbors in the same plane via purely elastic harmonic springs with a stiffness constant $k_h$.
By construction, these intra-plane bonds are considered unbreakable, thereby ensuring the structural integrity of each individual layer throughout the process.
Each particle on these planes is connected through an unbreakable spring of stiffness $k_0$ to a driving gauge that will control the shear strain applied to the system (see Fig. \ref{sketch}). This is our simplified consideration of the 3d bulk material surrounding the two planes.

The interaction  between the two planes is mediated by a set of inter-plane springs with elastic constant $k_v$ and rest length $l_0$, which is 
taken as the starting separation between particles. These springs connect nearest particles between the two planes, and therefore there are three springs 
attached to each particle.
The crucial point of the dynamics is the possibility for these inter-layer springs to break. Specifically, we proceed as follows. 
An intact spring exerts a force of intensity $f=k_v(l-l_0)$ along the stretching direction, which can be attractive or repulsive, where $l$ is the actual length of the spring. Yet, as soon as $l$ exceeds a prescribed critical length $l_c$, the 
spring ``breaks", and the force it exerts is irreversibly set to zero.
A set of neighbor springs failing sequentially will be the indication of a crack propagating in the system.

The mechanical state of the system is governed by its total potential energy $U$, which accounts for the three types of elastic interactions described in the previous paragraph. 
We define the coordinates on the planes as $x$ and $y$, and the perpendicular one as $z$.
For simplicity we will consider that the $z$ coordinates of the particles remain fixed, at the value in which the length of inter-layer springs is $l_0$.
Therefore, our main variables are $\mathbf{r}_{i}^{\alpha} = (x_{i}^{\alpha}, y_{i}^{\alpha})$, namely the position vector of site $i$ on plane $\alpha$ (where $\alpha \in \{1, 2\}$ represents the bottom and top planes, respectively). The total potential energy can be written as:

\begin{equation}
U = U_{\text{intra}} + U_{\text{inter}} + U_{\text{drive}}
\end{equation}
where the individual contributions are defined as follows:
\begin{equation}
U_{\text{intra}} = \frac{1}{2} k_h \sum_{\alpha=1}^{2} \sum_{\langle i,j \rangle} \left( \mathbf{r}_{i}^{\alpha} - \mathbf{r}_{j}^{\alpha} \right)^2\label{2}
\end{equation}
\begin{equation}
U_{\text{inter}} = \frac{1}{2} k_v \sum_{i,j(i)} \theta_{i,j(i)} \left( \left |\mathbf{r}_{i}^{(1)} - \mathbf{r}_{j(i)}^{(2)} \right |-l_0\right   )^2\label{3}
\end{equation}
\begin{equation}
U_{\text{drive}} = \frac{1}{2} k_0 \sum_{i} \left(  \mathbf{u}_{i}^{(2)}- {\gamma}\hat x \right)^2
+\frac{1}{2} k_0 \sum_{i} \left(  \mathbf{u}_{i}^{(1)}+ {\gamma}\hat x \right)^2
\end{equation}
In Eq. (\ref{2}) $\langle i,j \rangle$ denotes summation over nearest neighbors within the same plane. 
In Eq. (\ref{3}), given a position $i$ on plane (1), $j(i)$ indicates the three neighbors $j$ on plane (2) connected to $i$ through one of the $k_v$ springs. 
The dynamic state variable $\theta_{i,j(i)} \in \{0, 1\}$ explicitly tracks the irreversible damage history of the interface: $\theta_{i,j(i)} = 1$ if the bond is intact, and permanently switches to $\theta_{i,j(i)} = 0$ as soon as $|\mathbf{r}_{i}^{(1)} - \mathbf{r}_{j(i)}^{(2)}| > l_c$ for the first time. 
In  the driving term $U_{\text{drive}}$,  $\mathbf {u}_i^{(\alpha)}$ are the displacements of $\mathbf {r}_i^{(\alpha)}$ with respect to the initial equilibrium positions. Note that this term has been written for an applied shear strain in the $\hat x$ direction. Generalization to an arbitrary direction is straightforward. For the two-dimensional geometry we are considering, $\gamma$ has actually units of length.

Since we are interested in a quasi-static loading  protocol $\gamma(t)$, inertial effects will be neglected, and we consider an overdamped dynamic evolution. The equation of motion for each point mass $i$ on plane $\alpha$ is given by:

\begin{equation}
\eta \frac{d\mathbf{r}_{i}^{\alpha}}{dt} = -\nabla_{\mathbf{r}_{i}^{\alpha}} U
\label{dyn}
\end{equation}
where $\eta$ is the effective damping coefficient representing viscous dissipation.

Eqs. (\ref{dyn}) are solved under a sufficiently slow variation of the applied strain $\gamma$. In all results to be presented we in fact verified that a further
reduction in the time variation of $\gamma$ does not modify the results.
 Notice that during this evolution one or more of the inter-plane springs can break.\footnote{The failure of inter-plane bonds violates the hypothesis of quasistatic dynamics. This is a key point in cases of brittle fracture, where bonds can break sequentially in a rapid manner, leading to the propagation of a crack at finite velocity, which is strongly influenced by the inertia of the particles \cite{freund1990dynamic}. In our case we are interested in fatigue processes in which only a small number of bonds break at a given time, and therefore we still consider the overdamped dynamics to be  a reasonable assumption.} 
The main protocol of strain evolution that we will consider is as follows:

-The shear strain is monotonically increased from zero until it reaches a maximum value, $\gamma_{\max}$.

-Upon reaching this threshold, strain is reversed quasi-statically until it reaches a symmetric minimum value, $-\gamma_{\max}$.

-From here the process is repeated sequentially between the two limits $\pm \gamma_{\max}$.

We can have a flavor of how cyclic fatigue occurs by considering the sketch depicted in Fig. \ref{sketch}.
In (a) the system is in its minimum energy configuration, at $\gamma=0$ external strain. Notice that a few inter-layer springs are assumed to be already broken, and therefore are not drawn. When a strain $\gamma$ is applied (as indicated by the arrows in panel (b)) the odd-numbered springs are stretched, while even-numbered springs are compressed. Among the former, spring (1) is the most heavily stretched due to stress concentration effect caused by previously broken springs, and will be the first to break, as indicated. 
After spring (1) breaks, spring (2) can prevent spring (3) from failing. Therefore at this applied strain only one single spring is broken.
Upon inversion of the applied strain, now spring (2) will be the most stretched and the one that breaks. In this idealized situation one single spring is broken at each reversion of the applied strain. We will see that this kind of mechanism is robust, and persists in less ideal configurations.

We ran simulations for a system of 10000 particles on each triangular plane, and periodic boundary conditions both in $x$ and $y$. Values of the spring constants are $k_h=k_v=1$, $k_0=0.1$. The inter-layer springs break when they reach a length $l_c=1.5$ (from now on the distance between nearest neighbors in the relaxed configuration is taken as the unit of length, i.e. $l_0\equiv 1$). 

In realistic processes of fatigue crack advance, the existence of inhomogeneities and stress concentration effect play a crucial role. 
In our model, if we consider the ideal, homogeneous triangular lattice we would not obtain realistic results. We introduce inhomogeneities in the system in two ways. First, each inter-layer spring has some probability to be broken already in the initial configuration, before the external strain is applied. This will break the extreme homogeneity of the system, producing some stochasticity in the stresses over intact springs. In all simulations to be presented we take this probability to be $20 \%$.  

In addition, it will be crucial to include some additional stress concentrator that acts as a seed for a growing crack. This is in fact known to be a rather unavoidable situation in real samples, and therefore the most common situation to be studied. We introduce an initial seed crack in the system by breaking the inter-layer springs in some pre-defined spatial region of the system. Typically we will use seed cracks in the form of a stripe crossing the whole system, or alternatively in the form of a circular region.


\section{Numerical Results}
~\label{sec:results}

\begin{figure}[h]
\includegraphics[width=6cm,clip=true]{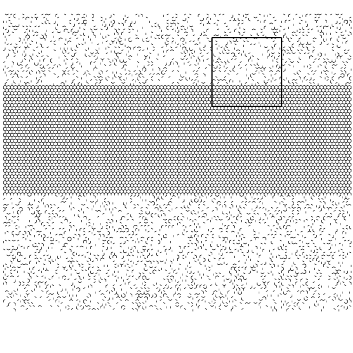}
\caption{The distribution of broken inter-plane links in the initial configuration (full system is depicted). Notice the stripe with all 
links broken, and the random distribution of a fraction of $20\%$ broken links in the rest of the sample. Square highlights the region that will be zoomed in Fig. \ref{snapshots}.
}
\label{links0}
\end{figure}

As a first configuration we consider a system with an initial crack in the form of a  stripe running along the $\hat x$ direction, and apply external strain along $\hat x$. Under these conditions the crack border is loaded in Mode III.
The choice of the value of the stripe width $W_0$ will be guided by the following considerations. 
Our aim is to observe the borders of the crack to advance upon cyclic application of shear strain, with minimum interference between the two. If $W_0$ is too small, we will have to overcome the tendency of the two borders to remain close to each other.  In the language of linear elastic fracture mechanics, we need to overcome the effect of the stress intensity factor at the crack tip. 
In our quasi-two-dimensional geometry, the stress intensity factor becomes constant once $W_0$ is larger than a quantity 
of the order of a screening length $\sim \sqrt{k_h/k_0}\simeq 3$ \footnote{Notice that at variance with our case, in the usual framework of a linear crack of length $L_0$ in an infinite two-dimensional medium\cite{freund1990dynamic}, the stress intensity factor is proportional to $L_0^{1/2}$, therefore producing a critical stress (the Griffith stress) that decreases as $L_0^{-1/2}$.}.
Therefore, for the moment we simply take a value of $W_0$ much larger than this length ($W_0=32$) for which the tendency of the two crack borders to remain bound is not observed. The spatial location of all broken links in the starting configuration is indicated in Fig. \ref{links0} (note that in that upper view, the links between all particles in the two planes define a honeycomb lattice).

\begin{figure}[h]
\includegraphics[width=9cm,clip=true]{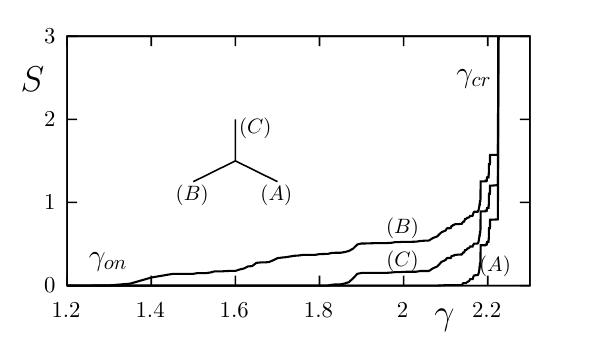}
\caption{Number of broken links $S$ as a function of the applied strain when this is monotonically increased from zero (normalization is such that $S=1$
represents the advance of each of the crack borders by one row of particles). Notice the progressive breaking of links, starting at $\gamma_{on}$ and up to the value $\gamma_{cr}$, where the system fails completely, and the fact that links of type B are those that break preferentially, as they are more stretched by the applied strain along the $\hat x$ direction.
}
\label{f1}
\end{figure}

We now proceed to the description of the numerical results.
First of all, as a reference, we consider the case in which the external strain is increased monotonically from zero, until the system fails completely (i.e., all inter-layer springs break). Due to the disorder and the initial crack present in the system this breaking does not occur in a single step. 
In Fig. \ref{f1} we show  the amount $S$ of broken inter-layer springs with different orientation as a function of the applied strain (we do not count in this number the bonds broken in the initial configuration). 
They are separated as types ``A", ``B", and  ``C", according to the spatial orientation they have. For a positive strain $\gamma$, B-bonds are stretched the most, C-bonds are less stretched, and A-bonds are compressed.  
Thus it is not surprising that for a given value of the strain, $S_{(B)}$ is the largest, followed by $S_{(C)}$, and  $S_{(A)}$ is the smallest. Notice also that once the value of  $S_{(A)}$ becomes different from zero, a small increase of $\gamma$ produces the complete failure of the system
at a critical (static) failure strain $\gamma_{cr}$ at which all links break at once (this is indicated by the vertical climb of the curves at $\gamma_{cr}$). 

\begin{figure}[h]
\includegraphics[width=9cm,clip=true]{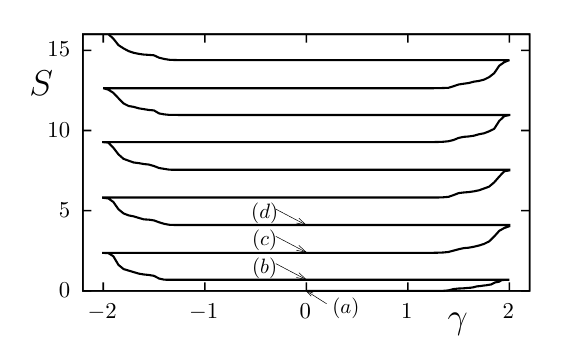}
\caption{Number of broken links $S$ (of any orientation) in the system when the applied strain is oscillated quasistatically between $\gamma_{\max}$ and $-\gamma_{\max}$.
Damage accumulates progressively as the number of cycles increases. Labels correspond to the snapshots in Fig. \ref{snapshots}.
}
\label{f2}
\end{figure}

\begin{figure}[h]
\includegraphics[width=10cm,clip=true]{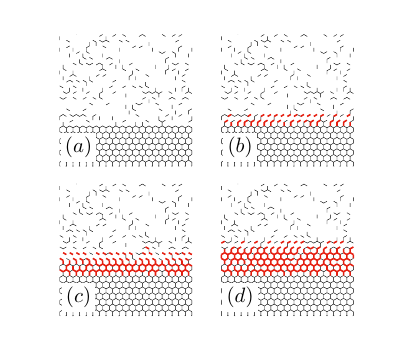}
\caption{The distribution of broken links within the small box depicted in  Fig. \ref{links0}. Panel (a) is the starting configuration, 
(b), (c), and (d) correspond to the tags in Fig. \ref{f2}.  Links broken by the applied strain are drawn thicker and red.
Notice how bonds of type B (A) are those mainly broken at positive (negative) applied strain.
}
\label{snapshots}
\end{figure}

Now we study the response of the system when strain is cycled between positive and negative values  $\pm\gamma_{\max}$, avoiding the failure of the system, namely $\gamma_{\max}<\gamma_{cr}$. The protocol is similar to the previous one, but when the predefined value $\gamma_{\max}$ is reached, strain is quasistatically reversed until $-\gamma_{\max}$, then back to $\gamma_{\max}$ and so on.
The main results are contained in Fig. \ref{f2}, and correspond to the value $\gamma_{\max}=2$. The amount of broken bonds per cycle rapidly reaches a stationary situation, indicating that we have a finite crack-tip advance per strain cycle. Snapshots highlighting the broken bonds at a few points of the process can be seen in Fig. \ref{snapshots}. This is the hallmark of a fatigue crack advance that is exclusively related to the cyclic nature of the applied strain, and which does not rely on the existence of plastic deformation in the material. 

\begin{figure}[h]
\includegraphics[width=9cm,clip=true]{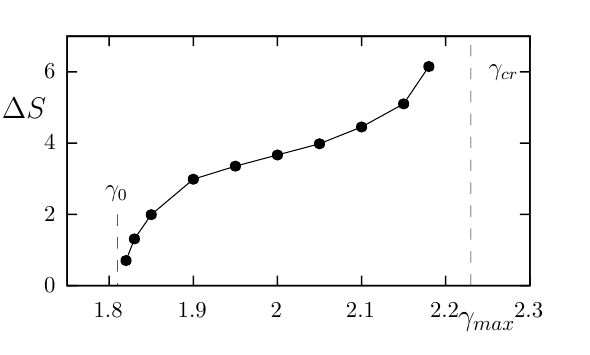}
\caption{The advance of $S$ per cycle as a function of the maximum applied strain $\gamma_{\max}$. $\Delta S$ increases sharply (pointing to a divergence) as $\gamma_{\max}\to\gamma_{cr}$. There is also an ``endurance limit" $\gamma_0$, such that if $\gamma_{\max}<\gamma_0$ the crack advance is eventually arrested.
}
\label{f6}
\end{figure}

The value of $S$ is strongly dependent on the value of $\gamma_{\max}$. Systematic simulations allow us to obtain the results presented in Fig. \ref{f6}. The value of $\Delta S$ (adavnce per cycle) is an increasing function of $\gamma_{\max}$, as expected. 
As $\gamma_{\max}\to\gamma_{cr}$, $\Delta S$ rises sharply, pointing to a divergence at $\gamma_{cr}$. In the other limiting case, as $\gamma_{\max}$ is reduced, we clearly observe that $\Delta S$ becomes zero if $\gamma_{\max}$ is sufficiently small. Although we do not have 
enough numerical data,  we conjecture that there is a sharp transition at $\gamma_0$, such that $\Delta S=0$ if $\gamma_{\max}<\gamma_0$. Our 
guess is guided by the similar behavior observed in fatigue of materials\cite{suresh1998}: the existence of a well 
defined $\gamma_0$ indicates the existence of an ``endurance limit" in our model, i.e., a minimum strain level below which the system can 
sustain theoretically an infinite number of cycles without failing. Also, notice that a behavior quite similar to the one presented in 
Fig. \ref{f6} was observed in models of depinning and yielding\cite{ferrero2025cyclic} in which 
some defining parameter of the model was subjected to quasistatic cycling. In that case the existence of an endurance limit was clearly 
established. Also it was claimed that the vanishing of $\Delta S$ as $\gamma_{\max}\to \gamma_0^+$ corresponds to a criticality similar to the 
one at the depinning point $\gamma_{cr}$. We conjecture that the same behavior occurs here, although more work will be 
necessary to put this claim on firmer grounds.


\begin{figure}[h]
\includegraphics[width=9cm,clip=true]{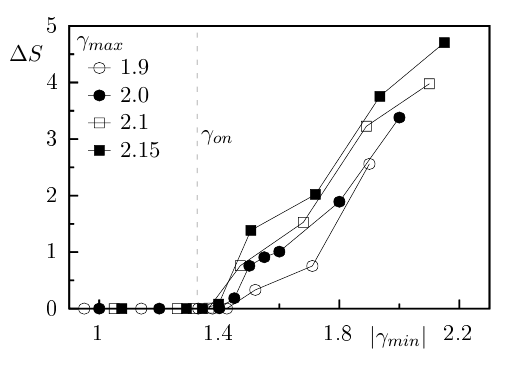}
\caption{Stationary advance per cycle $\Delta S$ as a function of the minimum strain value in the cycle. The advance reduces rapidly for asymmetric cycles. 
}
\label{deltas}
\end{figure}

The role of load reversal in the present fatigue mechanism has to be clearly indicated.
Systematic damage accumulation requires that the two extremes of the loading cycle bring
different subsets of bonds to failure, which in our model is achieved by reversing the
shear direction.
If, for instance, strain is
cycled between $\gamma_{\max}$ and 0, instead of between $\pm\gamma_{\max}$, no further
damage occurs after the first loading: since all interactions are elastic and the
evolution is quasistatic, once the bonds that fail at $\gamma_{\max}$ have broken, the
system returns to the same equilibrium configuration each time $\gamma_{\max}$ is
reached again, and no new bond reaches its breaking threshold. 
In order to visualize this effect, we ran simulations with a few different values of  $\gamma_{\max}$, and different values of $\gamma_{\min}$, ($\gamma_{\min}<0$, 
$|\gamma_{\min}|\leq \gamma_{\max}$). The results can be seen in Fig. \ref{deltas}.
For all the values of $\gamma_{\max}$ the fatigue advance completely disappears for $|\gamma_{\min}|$ near the value $\gamma_{on}$, which is the value at which the first links brake in the case of monotonic loading  (Fig. \ref{f1}). This is an indication that fatigue advance occurs only when each load inversion forces new fresh bonds to fail.

\section{Other loading modes and seed crack geometry}
~\label{sec:geometry}

\begin{figure}[h]
\includegraphics[width=9cm,clip=true]{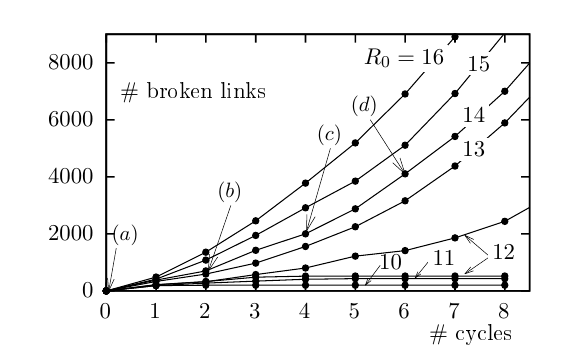}
\caption{Number of links broken by the fatigue process as a function of number of cycles applied, of amplitude $\gamma_{\max}=2.0$ for a sample with a circular initial crack of radius $R_0$ as indicated. Unbounded cyclic crack advance occurs when $R_0$ is larger than a critical value. 
}
\label{circular}
\end{figure}

\begin{figure}[h]
\includegraphics[width=8cm,clip=true]{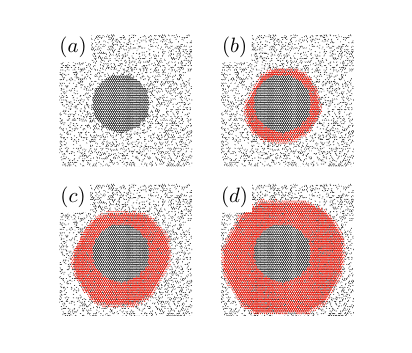}
\caption{Localization of the broken links at the points indicated in Fig. \ref{circular}, for an initial crack of radius $R_0=14$.
}
\label{snapshots2}
\end{figure}

In the previous Section a  pre-existent stripe-shaped crack was introduced in the system. 
In the present section we would like to comment on the generalization of  these results to other geometries.
To focus on a concrete case, we present the results for seed cracks of circular shape and different radii. 
Notice that in this case, and for strain applied along the $\hat x $ direction, the loading mode is in general mixed along the crack contour. Pure loading occurs only at the extreme horizontal  (Mode II) or vertical (Mode III) points of the contour. 
In addition, investigation of different seed crack radii will give information on the nucleation barrier for fatigue crack growth.
The main results in this respect are presented in Fig. \ref{circular}.
There we see how the number of links broken by the fatigue process evolves as a function of the number of cycles applied, for
initial circular cracks with different radii, under the same applied maximum strain ($\gamma_{\max}=2$) lower than the critical value determined in the previous section ($\gamma_{cr}\approx2.23$). There is a clear separation in the response of the system depending on the initial radius $R_0$. If $R_0$ is larger than some critical value ($R_0\gtrsim 12$), the evolution is consistent with the results of the previous section. In fact, the number of links broken per cycle follows the trend of the previous section if we take into account that here the total length of crack border increases linearly with the crack radius. 
In addition, a few snapshots of the system at different numbers of applied cycles (Fig. \ref{snapshots2}) show that the crack growth is fairly isotropic, beyond some stochasticity caused by the random broken bonds in the initial configuration. This is an indication that there is not any noticeable difference between growing under Mode II, or Mode III loading conditions.
When the initial crack radius is lower than the critical value, the evolution is halted after possibly the breaking of a few links, and the fatigue process is not able to progress. Notice in Fig.  \ref{circular} that for $R_0=12$ two different realizations produced an unlimited crack growth in one case, and a halted growth in the other.

We also mention the following possibility when the applied strain is larger than $\gamma_{cr}$. Although in the presence of a sufficiently large seed crack this would lead immediately to the total failure of the system, if a small seed crack is present, a single application of a supercritical strain may not lead to immediate failure. For instance, we have found that with a seed crack with $R_0=6$, the application of a strain $\gamma=2.3$ leads to a stationary fatigue growth until cycle \#6. Then the system fails completely at cycle \#7.

\section{Summary and conclusions}
\label{sec:discussion}

We have presented a minimal model showing that sub-critical fatigue crack
growth can occur in a system whose elements are elastic up to an
irreversible failure threshold, without plasticity, friction, or any
time-dependent process. The model represents a weak interface between two
elastic bodies loaded in shear. Progressive damage results from the
interplay of two key ingredients:  the fact that each bond fails more easily under one sense of the shear
than under the opposite one, and the reversal of the shear direction during
the cycle, which allows the two halves of each cycle to break different
subsets of bonds.

Under symmetric cycling below the static failure threshold
($\gamma_{\max}<\gamma_{cr}$), the crack advances by a well-defined amount
per cycle. This rate appears to vanish at a finite amplitude $\gamma_{\max}\sim\gamma_0$,
which plays the role of a fatigue threshold, and increases sharply as
$\gamma_{\max}$ approaches $\gamma_{cr}$ (Fig.~\ref{f6}). Asymmetric
cycles reveal a second condition. Growth stops when the reversed excursion
$|\gamma_{\min}|$ falls below $\gamma_{\rm on}$, the strain at which bonds
start to fail under monotonic loading (Fig.~\ref{f1}), and this
limit is nearly independent of $\gamma_{\max}$ (Fig.~\ref{deltas}). In
terms of the fraction of the forward strain that must be reversed, this
corresponds to 64--78\% for $0.85\lesssim\gamma_{\max}/\gamma_{cr}\lesssim0.96$.
Sustained fatigue thus requires $\gamma_{\max}\gtrsim\gamma_0$ and
$|\gamma_{\min}|\gtrsim\gamma_{\rm on}$ simultaneously. The physical content
of the second condition is simple: each load reversal must be strong enough
to break bonds that have not failed before. In the limiting case of a load
that does not change sign, the system shakes down after the first cycle and
damage stops altogether.

We have checked the robustness of these results with respect to several
details of the model. Varying the breaking length $l_c$ (in units of $l_0$)
at a fixed relative amplitude $\gamma_{\max}=0.9\,\gamma_{cr}(l_c)$, we find
steady crack advance for $1.3\le l_c\le 2.2$, with a rate per cycle that
decreases monotonically with $l_c$; the value $l_c=1.5$ used throughout lies
well inside this range. For $l_c=1.2$, crack growth is arrested after the
first cycle, because the bonds perpendicular to the shear direction (type C)
no longer break and remain behind the front as elastic bridges that shield
the crack tip. For larger $l_c$, the difference between the strains needed
to break a given bond under the two senses of the shear decreases, and
fatigue becomes restricted to amplitudes increasingly close to $\gamma_{cr}$:
for $l_c=3$, growth is negligible at $0.9\,\gamma_{cr}$ but clear at
$0.95\,\gamma_{cr}$.

Replacing the initially broken bonds by a random breaking length, uniformly
distributed between $l_0$ and $2l_0$, gives qualitatively the same
phenomenology: the weakest bonds fail during the first cycles, producing a
diffuse background damage, after which the crack front advances steadily.
A much broader distribution of breaking lengths, however, suppresses
fatigue except very close to $\gamma_{cr}$, because the strongest bonds
survive in the crack wake and again act as bridges. These observations point
to a general feature of the model: any intact element left behind the
advancing front shields the tip and, being purely elastic, is not degraded
by cycling. This is precisely the opposite of the situation in grain-bridging
ceramics, where cyclic fatigue originates in the frictional degradation of
such bridges~\cite{Lathabai1991,Ritchie1999,Dauskardt1993}; a model combining
both ingredients would be worth exploring. We also expect the mechanism not
to depend on the regular arrangement of the bonds in three particular
directions: what matters is that the bonds are not perpendicular to the
shear direction, so that positive and negative strains of the same magnitude
affect them differently. We refer to the Supplementary Material for a confirmation of this fact.

The need for a substantial reversal of the shear acting on the crack plane
sets the range of applicability of the mechanism. In the tension--tension
fatigue of structural ceramics the shear stress resolved on any plane does
not change sign, so the mechanism is inactive there, and the established
picture of crack-wake degradation~\cite{Ritchie1999,Ritchie2000} is not
affected. The mechanism could contribute, in addition to wake degradation,
only where the local shear reverses. This happens under fully reversed or
tension--compression loading~\cite{Reece1989}, and possibly in the presence
of residual stresses, for instance those caused by thermal expansion
anisotropy, which can shift the local loading cycle at grain boundaries.

The applicability domain of the present mechanism is
that of systems subjected to strongly reversed shear:
rolling and fretting contacts, components under reversed
torsion, and cemented granular materials and rock joints
under cyclic, typically symmetric, shear~\cite{Cerfontaine2018}. In a
sandstone, for instance, consolidated grains that are separated beyond some
critical distance detach and do not stick together again, which is exactly
what a breakable spring in our model represents, and discrete models based on
this picture are widely used~\cite{Potyondy2004}. 
 In this respect, we present as Supplementary Material results
on the fatigue advance effect in an atomic like three-dimensional system, with interaction 
between particles including a permanent repulsive part, and a ``breakable" attractive part, that align with main results of the paper.


It is instructive to contrast our results with recent studies of fatigue
failure in model glasses under cyclic
shear~\cite{Maity2026,Maity2026b,Khandare2026,Bhaumik2026}. These works have
uncovered a remarkably rich phenomenology. Below a fatigue limit, which
coincides with the yield strain amplitude, the system reaches a steady state
after a transient, whereas above it the number of cycles to failure diverges
as a power law upon approaching the fatigue limit. Damage is quantified there
through the accumulated plastic activity, which displays robust scaling
relations with the failure time, and failure occurs through the formation of
a system-spanning shear band. The situation addressed here is different in
several respects. In our model there are no plastic rearrangements: every
element behaves elastically until it breaks, and the only form of damage is
the loss of bonds. Failure proceeds through the advance of a pre-existing
crack front, rather than through the nucleation and growth of a shear band in
an initially homogeneous sample, and it occurs at amplitudes below the static
failure threshold $\gamma_{cr}$, i.e., it is a genuinely sub-critical
process. Both our study and the glass studies use fully reversed
cycles~\cite{Maity2026,Walker2026}. Asymmetric cycling would provide a
natural test to discriminate between the two scenarios. Our mechanism
requires $|\gamma_{\min}|\gtrsim\gamma_{\rm on}$, whereas damage driven by
plastic rearrangements, which leave residual stresses behind and display a
Bauschinger effect~\cite{Khandare2026}, is expected to persist for
$\gamma_{\min}\geq0$, as in the tension--tension fatigue of metals.

We thus regard the two scenarios as complementary limiting cases: fatigue
driven by accumulated plasticity, in systems where local rearrangements are
easily activated, and fatigue driven by sequential bond breaking at stress
concentrators, in systems where such rearrangements are suppressed. It is
suggestive in this respect that failure in well-annealed glasses becomes
increasingly abrupt and brittle-like~\cite{Bhaumik2026}, indicating that very
stable amorphous solids may be the systems in which the two pictures come
closest. In spite of their differences, both scenarios share the existence of
a sharp threshold separating arrested from progressive damage. If, as
conjectured in Sec.~\ref{sec:results}, the advance per cycle vanishes as a
power of $(\gamma_{\max}-\gamma_0)$, the number of cycles required to fail a
sample of finite size would diverge correspondingly as
$\gamma_{\max}\to\gamma_0^+$, offering a natural point of comparison with the
divergence of failure times observed in amorphous
solids~\cite{Maity2026,Bhaumik2026} and in cyclically driven depinning and
yielding models~\cite{ferrero2025cyclic}.

In conclusion, our minimal model establishes the conditions under which a
purely elastic--brittle system can undergo fatigue:  bonds whose failure depends on the sense of the shear, and a reversal
of the shear large enough for each half-cycle to break fresh bonds. These
conditions identify a plasticity-free route to cyclic damage, and point to
materials or interfaces under strongly reversed shear as the systems where it should be
sought.

{\bf Acknowledgements:} I thank Ezequiel Ferrero and Alfredo Caro for helpful discussions and insightful comments on the manuscript.

\bibliography{fatiga}

@book{suresh1998,
  author    = {Suresh, Subra},
  title     = {Fatigue of Materials},
  edition   = {2nd},
  publisher = {Cambridge University Press},
  year      = {1998},
  address   = {Cambridge}
}

@article{miller1982,
  author  = {Miller, K. J.},
  title   = {The short crack problem},
  journal = {Fatigue \& Fracture of Engineering Materials \& Structures},
  volume  = {5},
  number  = {3},
  pages   = {223--232},
  year    = {1982}
}

@inproceedings{forsyth1961,
  author    = {Forsyth, P. J. E.},
  title     = {A two stage process of fatigue crack growth},
  booktitle = {Proceedings of the Crack Propagation Symposium},
  year      = {1961},
  pages     = {76--94},
  address   = {Cranfield, UK}
}

@inproceedings{laird1967pathogenesis,
  author    = {Laird, C.},
  title     = {The pathogenesis of fatigue cracks},
  booktitle = {Fatigue Crack Propagation},
  series    = {ASTM STP 415},
  pages     = {131--180},
  year      = {1967},
  publisher = {ASTM}
}

@inproceedings{elber1971significance,
  author    = {Elber, Wolf},
  title     = {The significance of fatigue crack closure},
  booktitle = {Damage Tolerance in Aircraft Structures},
  series    = {ASTM STP 486},
  pages     = {230--242},
  year      = {1971},
  publisher = {ASTM}
}

@article{suresh1984propagation,
  author  = {Suresh, S. and Ritchie, R. O.},
  title   = {Propagation of short fatigue cracks},
  journal = {International Metals Reviews},
  volume  = {29},
  number  = {1},
  pages   = {445--475},
  year    = {1984}
}

@book{freund1990dynamic,
  author    = {Freund, L. B.},
  title     = {Dynamic Fracture Mechanics},
  publisher = {Cambridge University Press},
  address   = {Cambridge},
  year      = {1990}
}

@article{ferrero2025cyclic,
  author    = {Ferrero, Ezequiel E. and Jagla, Eduardo A.},
  title     = {Cyclic perturbations facilitate athermal creep in yield-stress materials},
  journal   = {Soft Matter},
  volume    = {21},
  number    = {31},
  pages     = {6307--6323},
  year      = {2025},
  publisher = {The Royal Society of Chemistry},
  doi       = {10.1039/D5SM00424A}
}

@article{Ritchie1988,
  author  = {Ritchie, R. O.},
  title   = {Mechanisms of fatigue crack propagation in metals, ceramics and composites: Role of crack tip shielding},
  journal = {Materials Science and Engineering A},
  volume  = {103},
  number  = {1},
  pages   = {15--28},
  year    = {1988}
}

@article{Ritchie1999,
  author  = {Ritchie, R. O.},
  title   = {Mechanisms of fatigue-crack propagation in ductile and brittle solids},
  journal = {International Journal of Fracture},
  volume  = {100},
  pages   = {55--83},
  year    = {1999}
}

@article{Ritchie2000,
  author  = {Ritchie, R. O. and Gilbert, C. J. and McNaney, J. M.},
  title   = {Mechanics and mechanisms of fatigue damage and crack growth in advanced materials},
  journal = {International Journal of Solids and Structures},
  volume  = {37},
  pages   = {311--329},
  year    = {2000}
}

@article{Dauskardt1987,
  author  = {Dauskardt, R. H. and Yu, W. and Ritchie, R. O.},
  title   = {Fatigue crack propagation in transformation-toughened zirconia ceramic},
  journal = {Journal of the American Ceramic Society},
  volume  = {70},
  pages   = {C248--C252},
  year    = {1987}
}

@article{Dauskardt1992,
  author  = {Dauskardt, R. H. and James, M. R. and Porter, J. R. and Ritchie, R. O.},
  title   = {Cyclic fatigue-crack growth in a {SiC}-whisker-reinforced alumina ceramic composite: Long- and small-crack behavior},
  journal = {Journal of the American Ceramic Society},
  volume  = {75},
  pages   = {759--771},
  year    = {1992}
}

@article{Dauskardt1993,
  author  = {Dauskardt, R. H.},
  title   = {A frictional-wear mechanism for fatigue-crack growth in grain bridging ceramics},
  journal = {Acta Metallurgica et Materialia},
  volume  = {41},
  pages   = {2765--2781},
  year    = {1993}
}

@article{Gilbert1995a,
  author  = {Gilbert, C. J. and Dauskardt, R. H. and Ritchie, R. O.},
  title   = {Behavior of cyclic fatigue cracks in monolithic silicon nitride},
  journal = {Journal of the American Ceramic Society},
  volume  = {78},
  pages   = {2291--2300},
  year    = {1995}
}

@article{Gilbert1995b,
  author  = {Gilbert, C. J. and Dauskardt, R. H. and Steinbrech, R. W. and Petrany, R. N. and Ritchie, R. O.},
  title   = {Cyclic fatigue in monolithic alumina: Mechanisms for crack advance promoted by frictional wear of grain bridges},
  journal = {Journal of Materials Science},
  volume  = {30},
  pages   = {643--654},
  year    = {1995}
}

@article{Gilbert1996,
  author  = {Gilbert, C. J. and Cao, J. J. and MoberlyChan, W. J. and De Jonghe, L. C. and Ritchie, R. O.},
  title   = {Cyclic fatigue and resistance-curve behavior of an in situ toughened silicon carbide with {Al-B-C} additions},
  journal = {Acta Materialia},
  volume  = {44},
  pages   = {3199--3214},
  year    = {1996}
}

@article{GavaldaDiaz2025,
  author  = {Gavalda-Diaz, O. and Saiz, E. and Chevalier, J. and Bouville, F.},
  title   = {Toughening of ceramics and ceramic composites through microstructure engineering: A review},
  journal = {International Materials Reviews},
  volume  = {70},
  pages   = {3--30},
  year    = {2025},
  doi     = {10.1177/09506608241308337}
}

@book{Johnson1985,
  author    = {Johnson, K. L.},
  title     = {Contact Mechanics},
  publisher = {Cambridge University Press},
  address   = {Cambridge},
  year      = {1985}
}

@article{Fiocco2013,
  author  = {Fiocco, D. and Foffi, G. and Sastry, S.},
  title   = {Oscillatory athermal quasistatic deformation of a model glass},
  journal = {Physical Review E},
  volume  = {88},
  pages   = {020301},
  year    = {2013}
}

@article{Leishangthem2017,
  author  = {Leishangthem, P. and Parmar, A. D. S. and Sastry, S.},
  title   = {The yielding transition in amorphous solids under oscillatory shear deformation},
  journal = {Nature Communications},
  volume  = {8},
  pages   = {14653},
  year    = {2017}
}

@article{Maity2026,
  author  = {Maity, S. and Bhaumik, H. and Athani, S. and Sastry, S.},
  journal = {Nature Physics},
  volume  = {22},
  pages   = {402},
  year    = {2026}
}

@misc{Maity2026b,
  author        = {Maity, S. and Khandare, P. and Bhaumik, H. and Sollich, P. and Sastry, S.},
  year          = {2026},
  eprint        = {arXiv:2602.21807},
  archivePrefix = {arXiv},
  primaryClass  = {cond-mat.stat-mech}
}

@misc{Khandare2026,
  author        = {Khandare, P. and Sastry, S.},
  year          = {2026},
  eprint        = {arXiv:2602.06174},
  archivePrefix = {arXiv},
  primaryClass  = {cond-mat.soft}
}

@misc{Bhaumik2026,
  author        = {Bhaumik, H. and Sastry, S.},
  year          = {2026},
  eprint        = {arXiv:2610.00655},
  archivePrefix = {arXiv},
  primaryClass  = {cond-mat.soft}
}

@article{thompson1956,
  author  = {Thompson, N. and Wadsworth, N. and Louat, N.},
  title   = {The origin of fatigue fracture in copper},
  journal = {Philosophical Magazine},
  volume  = {1},
  pages   = {113--126},
  year    = {1956}
}

@article{essmann1981,
  author  = {Essmann, U. and G{\"o}sele, U. and Mughrabi, H.},
  title   = {A model of extrusions and intrusions in fatigued metals {I}. {Point}-defect production and the growth of extrusions},
  journal = {Philosophical Magazine A},
  volume  = {44},
  pages   = {405--426},
  year    = {1981}
}

@article{mughrabi2009,
  author  = {Mughrabi, H.},
  title   = {Cyclic slip irreversibilities and the evolution of fatigue damage},
  journal = {Metallurgical and Materials Transactions B},
  volume  = {40},
  pages   = {431--453},
  year    = {2009}
}

@article{Potyondy2004,
  author  = {Potyondy, D. O. and Cundall, P. A.},
  title   = {A bonded-particle model for rock},
  journal = {International Journal of Rock Mechanics and Mining Sciences},
  volume  = {41},
  number  = {8},
  pages   = {1329--1364},
  year    = {2004}
}

@article{Cerfontaine2018,
  author  = {Cerfontaine, B. and Collin, F.},
  title   = {Cyclic and fatigue behaviour of rock materials: Review, interpretation and research perspectives},
  journal = {Rock Mechanics and Rock Engineering},
  volume  = {51},
  number  = {2},
  pages   = {391--414},
  year    = {2018}
}

@article{SornetteVanneste1992,
  author  = {Sornette, D. and Vanneste, C.},
  title   = {Dynamics and memory effects in rupture of thermal fuse networks},
  journal = {Phys. Rev. Lett.},
  volume  = {68},
  pages   = {612},
  year    = {1992},
  doi     = {10.1103/PhysRevLett.68.612}
}

@article{Kun2007,
  author  = {Kun, F. and Costa, M. H. A. S. and Costa Filho, R. N. and Andrade, J. S. and Soares, J. B. and Zapperi, S. and Herrmann, H. J.},
  title   = {Fatigue failure of disordered materials},
  journal = {J. Stat. Mech.: Theory Exp.},
  pages   = {P02003},
  year    = {2007},
  doi     = {10.1088/1742-5468/2007/02/P02003}
}

@article{Carmona2007,
  author  = {Carmona, H. A. and Kun, F. and Andrade, J. S. and Herrmann, H. J.},
  title   = {Computer simulation of fatigue under diametrical compression},
  journal = {Phys. Rev. E},
  volume  = {75},
  pages   = {046115},
  year    = {2007},
  doi     = {10.1103/PhysRevE.75.046115}
}

@article{Kun2008,
  author  = {Kun, F. and Carmona, H. A. and Andrade, J. S. and Herrmann, H. J.},
  title   = {Universality behind {Basquin's} law of fatigue},
  journal = {Phys. Rev. Lett.},
  volume  = {100},
  pages   = {094301},
  year    = {2008},
  doi     = {10.1103/PhysRevLett.100.094301}
}

@article{Vieira2008,
  author  = {Vieira, A. P. and Andrade, J. S. and Herrmann, H. J.},
  title   = {Subcritical crack growth: The microscopic origin of {Paris'} law},
  journal = {Phys. Rev. Lett.},
  volume  = {100},
  pages   = {195503},
  year    = {2008},
  doi     = {10.1103/PhysRevLett.100.195503}
}

@article{Oliveira2012,
  author  = {Oliveira, C. L. N. and Vieira, A. P. and Herrmann, H. J. and Andrade, J. S.},
  title   = {Subcritical fatigue in fuse networks},
  journal = {EPL},
  volume  = {100},
  pages   = {36006},
  year    = {2012},
  doi     = {10.1209/0295-5075/100/36006}
}

@article{Araujo2016,
  author  = {Ara{\'u}jo, M. S. and Vieira, A. P. and Andrade, J. S. and Herrmann, H. J.},
  title   = {Mesoscopic approach to subcritical fatigue crack growth},
  journal = {Phys. Rev. E},
  volume  = {94},
  pages   = {043003},
  year    = {2016},
  doi     = {10.1103/PhysRevE.94.043003}
}

@article{Roe2003,
  author  = {Roe, K. L. and Siegmund, T.},
  title   = {An irreversible cohesive zone model for interface fatigue crack growth simulation},
  journal = {Eng. Fract. Mech.},
  volume  = {70},
  number  = {2},
  pages   = {209--232},
  year    = {2003},
  doi     = {10.1016/S0013-7944(02)00034-6}
}

@article{Roux2000,
  author  = {Roux, S.},
  title   = {Thermally activated breakdown in the fiber-bundle model},
  journal = {Phys. Rev. E},
  volume  = {62},
  pages   = {6164},
  year    = {2000},
  doi     = {10.1103/PhysRevE.62.6164}
}

@article{Santucci2004,
  author  = {Santucci, S. and Vanel, L. and Ciliberto, S.},
  title   = {Subcritical statistics in rupture of fibrous materials: Experiments and model},
  journal = {Phys. Rev. Lett.},
  volume  = {93},
  pages   = {095505},
  year    = {2004},
  doi     = {10.1103/PhysRevLett.93.095505}
}

@article{GuozdenJagla2012,
  author        = {Guozden, T. M. and Jagla, E. A.},
  title         = {Fatigue crack propagation in a quasi one-dimensional elasto-plastic model},
  journal       = {Int. J. Solids Struct.},
  volume        = {49},
  pages         = {3224--3232},
  year          = {2012}
}

@article{Nguyen2001,
  author  = {Nguyen, O. and Repetto, E. A. and Ortiz, M. and Radovitzky, R. A.},
  title   = {A cohesive model of fatigue crack growth},
  journal = {Int. J. Fract.},
  volume  = {110},
  number  = {4},
  pages   = {351--369},
  year    = {2001}
}

@article{Lathabai1991,
  author  = {Lathabai, S. and R{\"o}del, J. and Lawn, B. R.},
  title   = {Cyclic fatigue from frictional degradation at bridging grains in alumina},
  journal = {J. Am. Ceram. Soc.},
  volume  = {74},
  pages   = {1340--1348},
  year    = {1991},
  doi     = {10.1111/j.1151-2916.1991.tb04109.x}
}

@article{AlonsoMarroquin2004,
  author  = {Alonso-Marroqu{\'\i}n, F. and Herrmann, H. J.},
  title   = {Ratcheting of granular materials},
  journal = {Phys. Rev. Lett.},
  volume  = {92},
  pages   = {054301},
  year    = {2004},
  doi     = {10.1103/PhysRevLett.92.054301}
}

@article{Zheng2022,
  author  = {Zheng, D. and Lin, S. and Ni, J. and Zhao, X.},
  title   = {Fracture and fatigue of entangled and unentangled polymer networks},
  journal = {Extreme Mech. Lett.},
  volume  = {51},
  pages   = {101608},
  year    = {2022},
  doi     = {10.1016/j.eml.2022.101608}
}

@article{Reece1989,
  author  = {Reece, M. J. and Guiu, F. and Sammur, M. F. R.},
  title   = {Cyclic fatigue crack propagation in alumina under direct tension--compression loading},
  journal = {J. Am. Ceram. Soc.},
  volume  = {72},
  number  = {2},
  pages   = {348--352},
  year    = {1989},
  doi     = {10.1111/j.1151-2916.1989.tb06133.x}
}

@article{Kida2012,
  author  = {Kida, K. and Honda, T. and Santos, E. C.},
  title   = {Mode {II} surface crack growth under rolling contact fatigue and cyclic shear stress in {Si$_3$N$_4$}},
  journal = {WIT Trans. Eng. Sci.},
  volume  = {76},
  pages   = {175--187},
  year    = {2012}
}

@article{Fujii2001,
  author  = {Fujii, Y. and Maeda, K. and Otsuka, A.},
  title   = {A new test method for mode {II} fatigue crack growth in hard materials},
  journal = {J. Soc. Mater. Sci. Japan},
  volume  = {50},
  number  = {10},
  pages   = {1108--1113},
  year    = {2001}
}

@article{Svetlizky2014,
  author  = {Svetlizky, I. and Fineberg, J.},
  title   = {Classical shear cracks drive the onset of dry frictional motion},
  journal = {Nature},
  volume  = {509},
  pages   = {205--208},
  year    = {2014},
  doi     = {10.1038/nature13202}
}

@book{HillsNowell1994,
  author    = {Hills, D. A. and Nowell, D.},
  title     = {Mechanics of Fretting Fatigue},
  publisher = {Kluwer Academic},
  address   = {Dordrecht},
  year      = {1994}
}

@misc{Walker2026,
  author        = {Walker, S. B. and Fielding, S. M.},
  year          = {2026},
  eprint        = {arXiv:2610.29615},
  archivePrefix = {arXiv}
}

@article{Bitzek2006,
  author  = {Bitzek, Erik and Koskinen, Pekka and G{\"a}hler, Franz and Moseler, Michael and Gumbsch, Peter},
  title   = {Structural Relaxation Made Simple},
  journal = {Physical Review Letters},
  volume  = {97},
  pages   = {170201},
  year    = {2006},
  doi     = {10.1103/PhysRevLett.97.170201}
}

@article{Guenole2020,
  author  = {Gu{\'e}nol{\'e}, Julien and N{\"o}hring, Wolfram G. and Vaid, Aviral and Houll{\'e}, Fr{\'e}d{\'e}ric and Xie, Zhuocheng and Prakash, Aruna and Bitzek, Erik},
  title   = {Assessment and optimization of the fast inertial relaxation engine ({FIRE}) for energy minimization in atomistic simulations and its implementation in {LAMMPS}},
  journal = {Computational Materials Science},
  volume  = {175},
  pages   = {109584},
  year    = {2020},
  doi     = {10.1016/j.commatsci.2020.109584}
}

@article{LeesEdwards1972,
  author  = {Lees, A. W. and Edwards, S. F.},
  title   = {The computer study of transport processes under extreme conditions},
  journal = {Journal of Physics C: Solid State Physics},
  volume  = {5},
  pages   = {1921},
  year    = {1972},
  doi     = {10.1088/0022-3719/5/15/006}
}


\section*{Supplementary Material: Fatigue crack growth in a three-dimensional particle model}

To test whether the mechanism discussed in the main text survives in a less
idealized setting, we performed quasistatic simulations of a three-dimensional,
disordered packing of cemented spheres. The model retains the two essential
ingredients of the lattice model: the elements are elastic up to an
irreversible breaking threshold, and there is no plasticity, friction, or
thermal activation.

\paragraph*{Model.}
We consider $N=7200$ spheres of equal radius $r_0$. Two particles at a distance $d$ interact through
\begin{equation}
V(d)=\epsilon\,(d-2r_0)^2/r_0^2,
\end{equation}
with $\epsilon$ the unit of energy. For non-cemented pairs this interaction acts
only for $d<2r_0$, so it is purely repulsive. Cementation is introduced on a
jammed configuration. Every pair with $d/r_0<2 \lambda $ $(\lambda>1)$ in that configuration is
bonded, and for bonded pairs $V(d)$ also acts for $2<d/r_0<2 \lambda $, providing
cohesion. A bond breaks irreversibly the first time its length exceeds
$2\lambda r_0 $. From then on the pair interacts only through the repulsive part,
and pairs that were not bonded initially never become bonded. The parameter
$\lambda$ plays the role of the breaking length $l_c$ of the lattice model. We
used $\lambda=1.25$.

\paragraph*{Sample preparation.}
Particles placed at random were brought to mechanical equilibrium at an
external pressure $P=6.7\times10^{-3}\,\epsilon/r_0^3$ by minimizing the enthalpy
with the FIRE algorithm~\cite{Bitzek2006,Guenole2020}, allowing a homogeneous
dilation of the box. The resulting packing is amorphous, with packing fraction
$\phi=0.654$ and mean contact number $z\simeq7.0$. After cementation, which
gives 11.2 bonds per particle, the system was relaxed again at the same
pressure. The cohesive forces produce a volume contraction of about 11\%, and
the final box is $L_x\times L_y\times L_z=(90.4\times15.1\times30.1)r_0^3$. Periodic
boundary conditions are used in $y$ and $z$. Bonds crossing the boundary in
$x$ were removed and a vacuum layer was added, so that the sample has two free
surfaces normal to $x$.

\paragraph*{Seed crack and loading.}
A simple shear strain $\gamma$ is applied in the $y$ direction through
Lees--Edwards boundary conditions in $z$~\cite{LeesEdwards1972}: the periodic
image of a particle at $(x,y,z)$ across the $z$ boundary is located at
$(x,\,y+\gamma L_z,\,z+L_z)$. An edge crack is introduced by removing all bonds
that cross the plane $z=L_z/2$ within a distance $a_0=22.6r_0$ (one quarter of
the sample length) from the left free surface. The crack front is therefore a
straight line parallel to $y$, and the applied slip loads it in Mode III, as
in the stripe configuration studied in the main text. The strain is varied in
increments $\Delta\gamma=2\times10^{-3}$, and after each increment the system
is brought to mechanical equilibrium, with residual forces below
$10^{-8}\,\epsilon/r_0$, with the box dimensions held fixed. The protocol is
strictly quasistatic and time-independent. We verified on a closely related
system that relaxing with overdamped dynamics instead of FIRE yields the same
stress--strain response up to the first bond-breaking avalanches, and only
minor quantitative differences afterwards.
\paragraph*{Results.}
The sample was first subjected to a monotonic deformation up to failure (Fig. \ref{fig:MD}(a)), dotted line.
The maximum of shear stress $\tau$ during this process is $\tau\simeq 0.12$, and occurs at $\gamma\simeq 0.37$.
Full breaking of the sample occurs at $\gamma_{cr}\simeq 0.47$.

The sample was then subjected to symmetric cycles between $\gamma_{\max}$ and
$-\gamma_{\max}$. 
The  evolution depends
sharply on the amplitude.
 For $\gamma_{\max}\lesssim 0.35$ the number of bonds broken per cycle decays
rapidly (577, 70, 34, 0),  and the stress--strain loop becomes stationary. The
system shakes down, as in the lattice model below the fatigue threshold.

Situation changes at $\gamma_{\max}=0.37$ ($\gamma_{\max}/\gamma_{cr}\simeq 0.79 $) (Figure~\ref{fig:MD}). 
The first reversal of the shear breaks a
large number of bonds. Part of them lie ahead of the crack tip and the rest
are distributed over the whole ligament. 
The fatigue crack is not a perfectly planar cut. Instead, it is a band of broken
bonds about two particle diameters thick, which wanders by a few particle radii
around the seeded plane. 
To characterize crack advance, we track the position of the active zone, defined as
the median position of the bonds broken between consecutive passages through
$\gamma=0$.
 After the first cycle the damage concentrates
in a narrow active zone, about $6\,r_0$ wide, located at the crack tip. This zone
advances at an essentially constant rate of about $6\,r_0$ per half cycle, with
100--200 bonds broken in each half cycle (Fig.~\ref{fig:MD}(c)). The stress
amplitude decreases steadily as the crack grows. The crack crosses the whole
ligament, and the sample fails during the sixth cycle, although it had
withstood the first loadings to the same strain.


\setcounter{figure}{0}\renewcommand{\thefigure}{S\arabic{figure}}
\begin{figure}[t]
\centering
 \includegraphics[width=8cm]{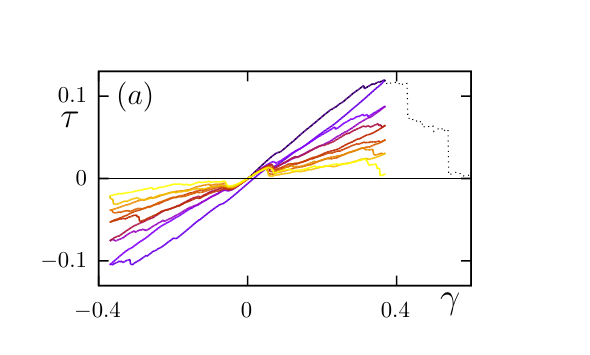}
 \includegraphics[width=8cm]{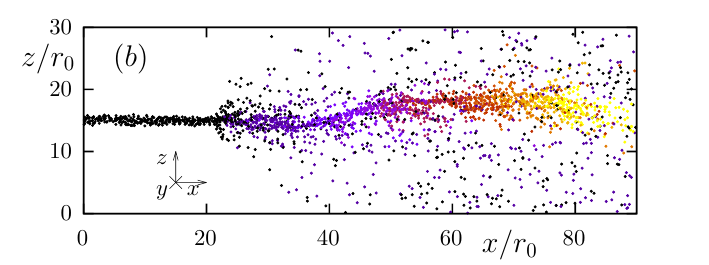}
 \includegraphics[width=8cm]{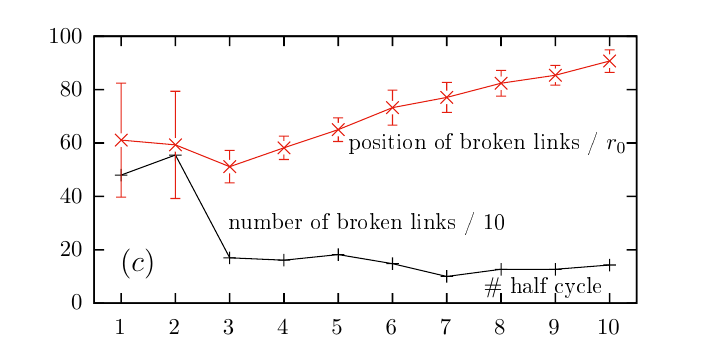}
\caption{Cyclic shear of a cemented packing of 7200 spheres ($\lambda=1.25$)
with an edge crack of length $a_0=22.6\,r_0$, for $\gamma_{\max}=0.37$.
(a)~Stress/strain curve. Dotted line is the result of a monotononic loading until failure.
Continuous lines are the result of successive cycling between $\gamma=\pm 0.37$.
(b)~Spatial position of broken bonds, and its time of occurrence [colors correlate with those in the stress/strain curve at (a)]. Notice the existence of diffuse damage during the first two half cycles, followed by the advance of a fatigue crack in subsequent cycles. The seed crack corresponds to the broken bonds at $x< a_0$.
(c)~Number of broken links per half-cycle, and position of the active damage zone: median, and 25--75\% range of the $x$
coordinate of the bonds broken between consecutive passages through
$\gamma=0$. }
\label{fig:MD}
\end{figure}

These results reproduce the main features of the lattice model. Below a
threshold amplitude, which here lies between 0.35 and 0.37, damage arrests
after a few cycles. Above it, the crack grows cycle after cycle by the
sequential breaking of bonds that are stretched by alternate senses of the
shear. Bonds oriented close to $\pm45^\circ$ in the $yz$ plane are the most
strongly stretched by one sense of the shear and compressed by the other, and
they are the ones that break preferentially near the crack tip. Bonds nearly
perpendicular to the slip direction are stretched only weakly in both half
cycles and tend to survive behind the front. Consistently with the role of
such bridging elements discussed in the main text, we find that the range of
cohesion matters. For $\lambda=1.2$ no sustained crack growth was observed in
any of the geometries studied: damage always arrested after one to three
cycles, leaving poorly oriented bonds intact behind the front. For
$\lambda=1.25$ and $1.3$, fatigue crack growth was observed.

\end{document}